# How are your robot friends doing? A design exploration of graphical techniques supporting awareness of robot team members in teleoperation

Stela H. Seo    James E. Young    Pourang Irani

Department of Computer Science
Faculty of Science
University of Manitoba
{stela.seo, young, irani}@cs.umanitoba.ca

**Abstract** While teleoperated robots continue to proliferate in domains including search and rescue, field exploration, or the military, human error remains a primary cause for accidents or mistakes. One challenge is that teleoperating a remote robot is cognitively taxing as the operator needs to understand the robot's state and monitor all its sensor data. In a multi-robot team, an operator needs to additionally monitor other robots' progress, states, notifications, errors, and so on to maintain team cohesion. One strategy for supporting the operator to comprehend this information is to improve teleoperation interface designs to carefully present data. We present a set of prototypes that simplify complex team robot states and actions, with an aim to help the operator to understand information from the robots easily and quickly. We conduct a series of pilot studies to explore a range of design parameters used in our prototypes (text, icon, facial expression, use of color, animation, and number of team robots), and develop a set of guidelines for graphically representing team robot states in the remote team teleoperation.

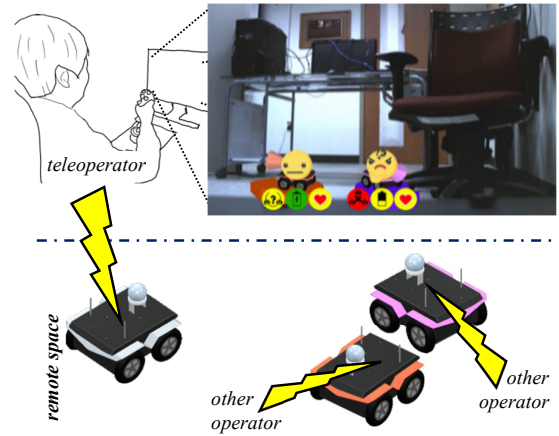

**Fig. 1** An operator teleoperates a robot located in remote environments with two team robots. While they are controlling one robot using a joystick, they maintain awareness of team robots using the robot state representations (screen bottom). The representation shows team robot states using icons and emotional information encoding.

## 1 Introduction

Teleoperated robots are becoming increasingly common and affordable, being used in any situation where it can be difficult, dangerous, or expensive to send people. This includes search and rescue, military reconnaissance, exploration (e.g., deep sea), or industrial equipment inspection. Multi-robot teleoperation gains more attention in these field as well [4,39]. While using multiple robots brings benefits including productive and effective task performance, an operator suffers increasing cognitive load as more robots are added to the team. Teleoperating multiple robots increases the chance of human errors; user error remains a primary cause of mistakes in teleoperation scenarios [3,12].

In team teleoperation scenarios whether robots are autonomous or controlled by teammates, an operator must maintain team robots' states for efficient team cohesion and seamless task transitions between robots (e.g., making decisions for robots). There is a great deal of information from the remote robots that the operator must monitor and understand, including a range of specialized sensors and video and audio feeds. Operators need to develop and maintain the situation awareness of the robots, their states, and their interaction with the environment [26], to make decisions and operate the robots in real-time. Among the myriad of challenges in team teleoperation scenarios, one is the teleoperation interface: input and output hardware, displays, widgets, and controls need to be designed to support operators in controlling robots, monitoring their states, and ultimately in making informed decisions in real-time.

In this work, we focus on how to help an operator to maintain awareness of team robot states. We explore a set of prototypes with different interface parameters to simplify complex robotic team member states, aiming to be easy for an operator to understand them. We choose a set of robot states which are generally applicable to various mobile robots. Our prototypes consist of different information representations including text, icon, and emotion representations, and graphical visualization parameters including color, animation, and number of team robots. We describe our reasons for choosing the robotic states, representations, and parameters.

We developed a mock team teleoperation scenario, task, and environment. We directed our scenario to fully support our design exploration. In our scenario, an operator is working on their robot manually in an environment where team communication is limited (e.g., due to the language barrier or limited communication bandwidth). In addition, the operator must maintain their awareness of team robot states using different state representations. We developed on-screen questionnaires to periodically test people's awareness of team robot states.

We conducted a series of pilot studies, iteratively exploring our robot state representations while investigating people's thoughts on the interfaces throughout the process. Our goal is to iteratively collect people's opinions on various team robot state representation designs and prepare an initial design guideline for graphically representing team members' robotic states in teleoperation interfaces in an ecologically valid study scenario. For this reason, in the study, our participants actually teleoperated a mobile robot located in a different room in real-time while we tested how they maintained their awareness of team robots with our various prototypes.

We share the lessons we learned throughout this process and discuss advantages and disadvantages of our team robot state representations, developing guidelines to help interface designers to make informed decisions for designing team robot state representations in teleoperation interfaces. In addition, we include a reflection section of our evaluation methods with recommendations for ongoing work to describe challenges we faced. Our work contributes to the larger and ongoing discussions on teleoperation interface design elements and shape future directions for in-depth exploration.

## 2 Related Work

In teleoperation, the user interface is a crucial factor for an operator's task performance [5,28,33] and situation awareness [9,26,33], as the user interface is the only gateway that connects an operator and the remote teleoperated robots. Improving interface usability is an ongoing challenge for researchers to reduce human operator error for teleoperation problems [3,12], to use human resources efficiently [22], and to reduce operator's cognitive load [28,33].

Many researchers worked on improving the level of automation [6,30], control mechanism of teleoperated robots [2,14,19,23], and visual interface [28,33]. However, we still do not fully understand how graphical design elements can be applied to multi-robot teleoperation interfaces and how they impact an operator's awareness of their team robot members. Especially, we do not know how to leverage social techniques (e.g., facial expressions) to provide useful information to an operator [29]. We explore a set of design elements including emotional information encoding to understand their impacts on teleoperation task performance and user perception.

Interface designers and researchers have attempted to reduce the operator's cognitive load in many ways, including improving the robot's artificial intelligence [30], reducing complex data to coarse-grained [34], high-level information visualization more relevant to a task [38], and drawing attention to important errors to reduce the possibility of missing an important event [28]. Our work follows this common goal: we design graphical overviews for high-level team robot states with various design elements and search for ways to potentially reduce the operator's cognitive load.

In previous work in human-agent interaction, researchers found that good team work involves preemptive actions to help team members [7] and positive engagements between team members [27]. In order to support these actions, the system must help an operator (a team member) to be aware of other team members (i.e., helping a person to maintain high situation awareness). We pursue the same high-level goal: we want an operator to maintain high-level situation awareness. In other words, we explore how design elements can impact people's high-level awareness of team robot members in a cognitively taxing team teleoperation scenario.

In human-computer interaction and human-robot interaction, researchers have explored icon representations [1,11,31] and emotional information encodings [25,32,35] in various applications. While icons and emotional information encodings can help people's situation awareness, they are not studied for multi-robot teleoperation interface designs. Hence, we build our state representations with these techniques and explore their impacts.

Many teleoperation interfaces are built around the main video source. In video-centric interfaces, designers try to reduce interface occlusions [33] or simplify as much information as possible [34]. We follow the theme by having small robot state representations with the high degree of information.

## 3 Graphical Approaches for Representing Robotic States of Team Members

In our exploration, an operator (a user) is a member of a teleoperation team (**Fig. 1**) and maintains awareness of team robot's states to improve overall teamwork efficiency.

We built a video-centric teleoperation interface, which has a robot's video feed as the main remote information source. We added our graphical team member state representations to the interface for users to maintain their awareness of team members with information representations and graphical parameters. We worked closely with a local artist from ZenFri Inc. to create visually aesthetic assets.

### 3.1 Team Robot States

In our exploration, a user teleoperates a robot in real-time and performs a task. Simultaneously, they monitor activities and robotic states of other team members including their movement (direction and wheel rotation), view direction, and different internal states.

A robot has many internal states but not every state is equally important for a task. With these in mind, we choose four internal states which are generally available on many mobile robots but specific enough to know their functionalities: connectivity, battery level, physical damage, and message availability. For high-level awareness purposes, we abstract the details out to three levels (**Table 1**).

Connectivity state can be important for mobile robots when the remote environment is unstable (e.g., search and rescue) to know how responsive or reliable a robot is. Disconnected robots can block an important route for other robots.

Many mobile robots have a battery to operate in the remote environment. Having low battery can signify an operator to return the robot to the base for recharging or for a team member to replace the member's role in advance.

As robots are working in a potentially dangerous environment, they may receive physical damage from their activities. This state becomes more important in search and rescue, as a damaged robot may not function well in an important situation or may drag down the team's workflow and efficiency.

Mission feedback and messages can provide an important information to team members depending on the message type, especially when the team has limited communication capabilities (e.g., an international operation team who are physically distant or have disabled radio communication).

In addition, we represent the robot's current movement, whether it is moving or not, and which direction it is facing and looking at.

### 3.2 Incorporating Robot States on Screen

We created a composite indicator that includes general state information for each team robot. This composite indicator has an avatar and an indicator for each state on top (**Fig. 2**).

The robot avatar is simply a 3D model of the robot. It is animated to show the robot's turning or moving (the wheels move, and the robot rotates). The avatar is placed on a disc which has an arrow tip indicating the moving direction, and a cone in front of the camera indicating its look direction. The 3D model animates and moves as the team robot moves.

We explored various information representations and graphical visualization parameters: icons are often used to simplify a complicated message, emotional encoding can simplify a message using people's social skills, color can be used to differentiate the importance of states (e.g., urgent, normal), animation can draw people's attention. The number of robots can explain the difficulty of maintaining awareness of team robot states, despite our simplification. Using these representations and graphical visualization parameters, the team robot's four states are overlain on top of the robot model.

### 3.3 Information Representation

We selected three major information representation techniques for our exploration: text, icons, or emotional information encoding.

#### 3.3.1 Text Team Robot State Representation

We developed a text state representation to use as a base case in our exploration. We selected text as it is quick and simple to develop and may be universally understandable. We expect the text to be easily understandable but not to be preferred over icon or emoji state representations.

Each state is formed with two words (one descriptive adjective and one noun, **Fig. 3**). To keep consistent size (resolution) with other state representations, we had to tilt the text about 45 degrees (icon is square while text is rectangle if not rotated). We added shadow to enhance visibility and readability.

Table 1. Four general robot states in our exploration. We abstract the details out to three categories.

|                  |                     |                   |                   |
| ---------------: | :-----------------: | :---------------: | :---------------: |
| **connectivity**     | strong connectivity | okay connectivity | weak connectivity |
| **battery level**    | strong battery      | okay battery      | weak battery      |
| **physical damage**  | no damage           | light damage      | heavy damage      |
| **mission message**  | no message          | message           | urgent message    |

### 3.3.2 Icon Information Encoding

Icons can metaphorically convey a complicated message [11] and can be quickly understandable at a distance [18] and while moving [20]. Well-designed icons can quickly convey its meaning to people. We would like to confirm if previous findings of icons are also applied in teleoperation interface by comparing our icon team robot state representations to our base case (text representation). We expect the icon to be comparably understandable with the text but appreciated aesthetically. In addition, icons may help people to increase their overall awareness by quickly conveying its meaning.

Icons are matched by to the text state information (they have one-to-one mappings, **Fig. 4**). In the designing process, we tried to create universally understandable icons, such as connection between robots using lines; filled, partially filled, or empty dry battery; heart (often expressing health) with different sizes; and empty, question, and exclamation marks. We iterated on our icon design multiple times. For example, because the lines on connectivity icons were difficult to distinguish, we added an additional mark (check, question, or x marks) in the middle.

### 3.3.3 Emotional Information Encoding

We developed facial expressions, which may help people to understand complex and difficult-to-understand information easily. In human-robot interaction, researchers found that aggregated facial expression "may well serve as both symptoms of an underlying state and communicative signals" [15,24]. In addition, social expression can increase communication bandwidth [35,36]. With a team robot's facial expressions, we expect people to take a quick glance at the representation, build general idea of the robot's current states, and further increase their overall awareness.

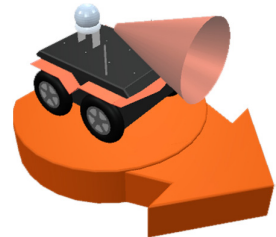

**Fig. 2** A team robot indicator with its moving direction (arrow on the bottom) and look direction (cone on the top).

Our team robot facial expressions (i.e., emojis on team robots, **Fig. 5**) aggregate all states into a single facial expression. For example, if all states are positive, the robot shows a happy emotion. For emojis, unlike text and icon representations, we used two colors: red (angry) for overall negative states and yellow (all other emotions) for overall positive or neutral states. We did not include green because a green face is perceived as a sick face.

### 3.4 Graphical Visualization Parameters

Along with the information representations (text, icon, and emoji), we explore graphical parameters to understand their benefits for visualizing team robot states. We have selected three parameters: color coding, animation, and a number of team robots.

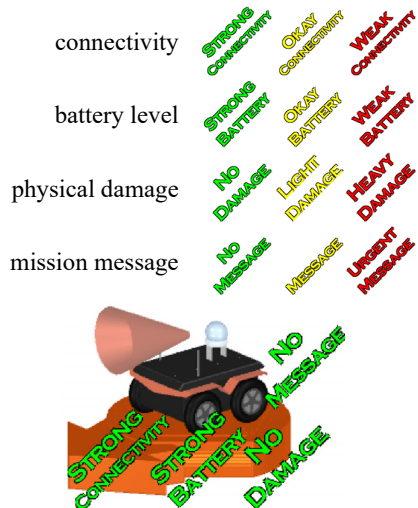

**Fig. 3** Text representations. Each state consists of two words: one descriptive adjective and one noun for a state. We added shadow to enhance visibility and readability of words when they overlay on the team robot. An example is shown on the bottom.

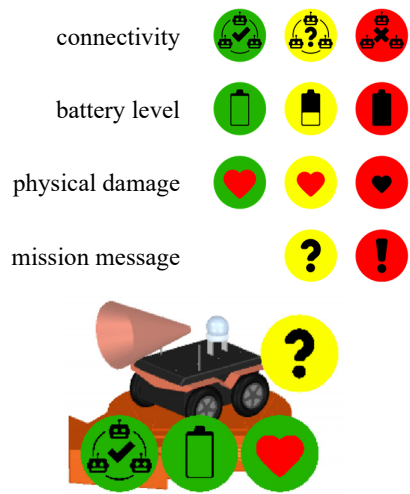

**Fig. 4** Icon representations for each robotic state. They have one-to-one mappings to the text. The icons are placed around the robot model, in the same position as the text indicators.

### 3.4.1 Color Coding of Robot States

We included color as a parameter to explore how it impacts people's cognitive load in maintaining awareness of team robot states. We prepared two versions: color and grayscale (**Fig. 6**). In color version, a positive state is mapped to green, a negative to red, and an intermediate to yellow.

Color coding can enhance an interface's effectiveness [21], for example, showing levels of states, expressing layers of a circuit board, or enhance visibility with color contrast. This can be useful to visualize a level of robot states (e.g., severity or urgency). On the other hand, grayscale provides a simple view and less visual noise overall.

We expect grayscale versions to be hard to distinguish different states compared to color versions; we are interested in learning about people's thoughts and experiences with grayscale versions as well.

### 3.4.2 Animated and Non-animated Icons

We added an extra dimension on our state indicators: animated or static. Animation can draw attention from people, the desired result for the urgent information; however, it can be distracting and may impact an operator's performance negatively [28]. As such, we feel it is important to investigate both static (not distracting but may ignored) and animated (may distracting but hard to ignore) variants.

We explore how animation impacts people's awareness of team robot members. We choose the edges on the animation scale: constantly animating or statically presenting, but localized (i.e., only animated in the boundary) and not dramatic. There is no transition between one level to another. Two example s are shown in **Fig. 7**.

### 3.4.3 The Number of Team Robots

As the number of team robots may impact people's task performance as well as awareness of team robot states, we compared two cases: one team robot and two team robots.

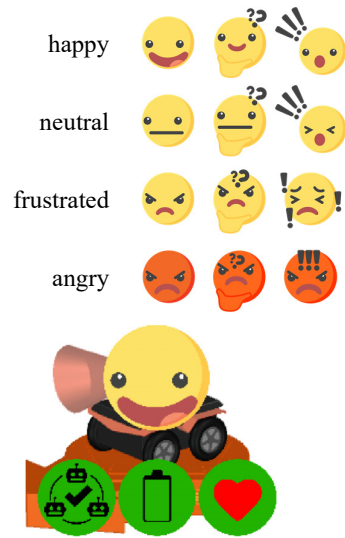

**Fig. 5** Emoji shows the condition of general states using a facial expression. The message state is embedded onto facial expressions – the left face on each category is for no message, the middle for a message, and the right for an urgent. An example on the bottom shows that a team robot has all positive states and no messages.

In one team robot condition, we assume a team of two robots: one is controlled by the user and another by another operator. In two team robot condition, we assume a team of three robots: one is controlled by the user and two by other operators. To prevent confusion, we refer these cases to the number of team robots without the robot controlled by the user (e.g., one team robot vs. two team robots), instead of the total numbers of robots in the team.

We color coded team robots. The first team robot (on left if two are present) has an orange body and the second is on the right side of the first and has a purple body. In the one team robot case, the orange team robot (**Fig. 8**, left) would be shown while the purple robot (**Fig. 8**, right) would not be shown.

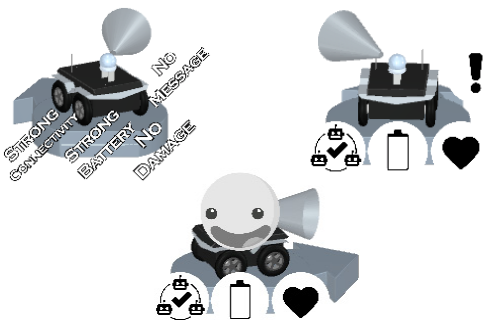

**Fig. 6** Every asset in the grayscale version: text, icon, and emoji representations. In the grayscale version, the team robot is also in grayscale. Examples with each representation shows on the right.

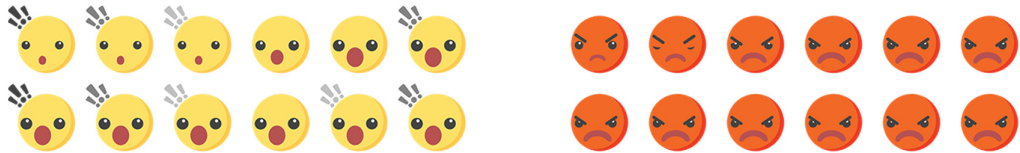

**Fig. 7** Two sprites (happy with an urgent message and angry) – we only present every third frame here due to the space constraint. The original sprite has 40 frames which provide smooth animation. One cycle of the animation takes about 1.33 seconds. The animation constantly plays in our animated representations.

## 4 Exploratory Study Design

From our related work, we have some expectations in how our designs may impact people's opinion (e.g., icon would be comparably understandable with the text); however, it is not clear if these previous findings would be applicable in egocentric teleoperation interfaces. Thus, we explore how state indicators with different information representations and visualization parameters impact people's awareness of team robot states in teleoperation scenarios.

One of the challenges in our investigation is to evaluate various prototypes in teleoperation. We need a general scenario and task, which allow us to investigate our research questions with a high ecological validity.

### 4.1 Teleoperation Scenario Test Bed

We developed a test bed which lets us evaluate our prototypes. Our driving principle is the high ecological validity – our participants teleoperate a robot as like a real-world teleoperator controls a robot with real-time video feeds instead of watching videos, navigate the environment with real obstacles, and perform a task actively, while maintaining awareness of constantly updating team robot states.

#### 4.1.1 Navigation Task

We chose a logistic task: a robot goes to a location to pick up an object and then moves to another to unload. It is a simple task yet happens in many applications (even in search and rescue) along with other tasks.

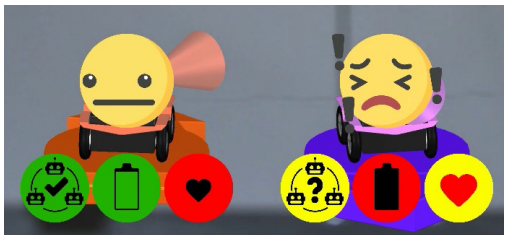

**Fig. 8** Our team state representations for two team robots with emotional information encoding. Note that the message state is embedded onto the face.

We setup the environment to have six designated locations and simple routes that connect these locations (**Fig. 9**). Although the routes are simple, we did not include any long straight paths. With this setting, a participant faces navigation challenges such as localization, orientation, and avoiding obstacles.

If the robot collides or pushes obstacles during the studies, an on-site researcher quickly fixes them without interrupting the robot, so that a participant always faces the same obstacle settings.

#### 4.1.2 Action Task

Once the robot arrives at a designated location, marked with a printed paper on the wall, the screen indicates a specific task, loading or unloading a cargo. Upon a completion of the task, the participant is asked to navigate to next designated location.

To complete the task, a participant must fill up a progress bar. We wanted to provide some challenge, while not increasing a participant's cognitive load. For example, manipulating a physical object would increase participants' workload. The progress bar fills up when a participant presses and releases a button on the joystick. The bar slowly loses its progres-sion if the button is not repeatedly pressed and released. This provides a small challenge and the feeling of achievement once it is completed.

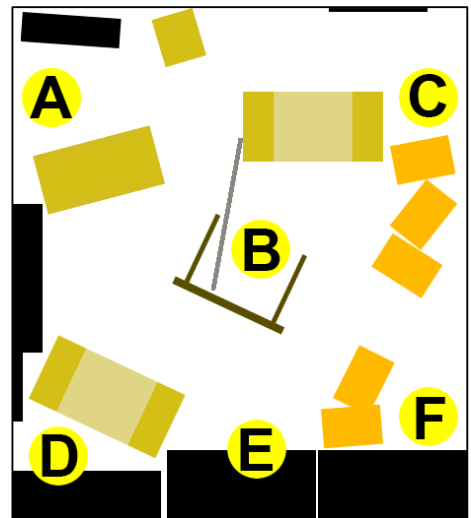

**Fig. 9** A top-down map of the environment. The map shows two underpasses (desks, lighter shading), obstacles, and designated locations. The map is always on the secondary monitor during studies, but the robot's current location is not marked.

Our system adjusts the filling rate based on the elapsed time on the location. If a participant spent a long time to fill the bar, the system increases the rate per click. We want a simple but active task which does not hinder our exploration of research questions, not causing frustration to the participant.

### 4.2 Real-time Teleoperation Interface

We developed our teleoperation interface using Unity3D. We placed the state representations on the video feed updating in real-time, and the mission objective at the top-center along with the progress bar which displays when the task is active (**Fig. 10**).

Since our goal is to explore how the information representations and the graphical visualization parameters impact people's team awareness, the consistency of team robots is important in the study. Thus, we created simulations of the team robots' movements (movement, look direction) and working states that a participant must pay attention to.

### 4.3 Measurements

We use on-screen self-report questionnaires to probe people's awareness of team robot states. Since our state inquiry questionnaire involves short-term memory, it is important to minimize the delay to get to questions. Further, as we are measuring their awareness, not ability of reading indicators, we blank the interface when questionnaires are up.

### 4.4 Apparatus

We use the Jackal UGV robot (**Fig. 11**, left), built by Clearpath Robotics Inc., which is 43cm wide, 51cm long, and 25cm tall, and about 17 kg. The robot is running on ROS indigo.

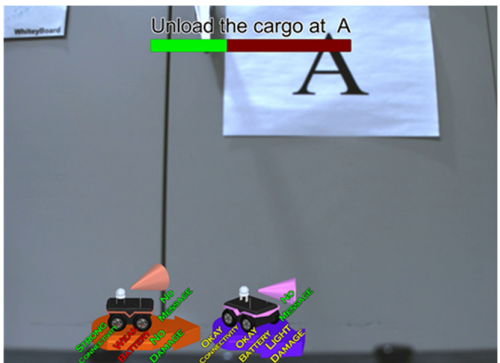

**Fig. 10** A participant is unloading the robot. The mission goal and its progression are shown at the top-center of the screen. Note that while the goal is always displayed, the progress bar only appears when the robot is at the designated location.

The robot has two cameras: Axis P5514-E PTZ Network Camera and Point Grey Flea3 FL3-U3-13E4C-C. At first, we used Axis camera for the remote video feed; however, after the first study, Axis camera was severely damaged from accidents (e.g., hitting the robot to obstacles). For the rest of studies, we used Point Grey Flea3.

We use a PC with two monitors, a 28-inch UHD monitor for our teleoperation interface and a 24-inch FHD monitor to show the map. A participant controls Jackal using a gaming joystick (jet fighter replica gaming joystick, **Fig. 11**, right). The monitor is about 65cm far from the participant, the keyboard and mouse are located on the keyboard tray under the desk and the joystick is placed on the desk; however, we allow the participant to change the initial desk setup for their convenience.

## 5 Design Exploration

We conducted a set of pilot studies. We take an explorative and iterative approach with small sample sizes but many variants. We focus on high-level results and lessons for future investigations.

We recruited our participants from the general university public. Every participant received $15 at the beginning of the study for their time. With people's comments collected in our studies, we conducted open coding approach to understand people's implications on our representations with different design elements. We introduce emerged themes in each section. Note that the participant ID is incremental but not sequential for different studies.

### 5.1 General Procedure

At the beginning of the study, we explain robot teleoperation and its potential use with real-world examples. With understanding of the problems, participants can realize the benefits of multiple robots (e.g., quickly searching for victims). We use team activities (e.g., sports) as examples and explain possible teamwork in teleoperation. Further, we explain challenges such as increasing cognitive load.

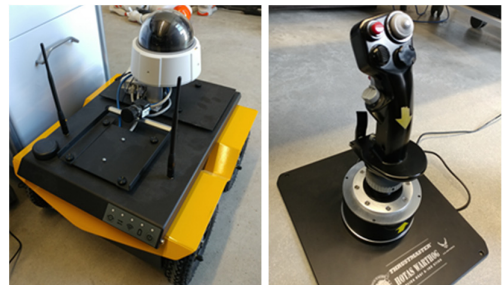

**Fig. 11** Jackal UGV robot from Clearpath Robotics Inc. (left) and a gaming joystick (right).

After the explanation, we conduct the demographics questionnaire and provide a training session when the person controls the robot using a joystick in real-time. In the training session, we introduced team robot states, our state representations, teleoperation tasks, and questionnaires. Before continuing, we provide enough time for the person to try different movements with our mobile robot. The researcher waits for the person to decide to move to the next phase of the study. If, however, they did not try a certain movement (e.g., going backward), the researcher requests they try it at least once.

We measure people's awareness of their team robot states using a questionnaire (state inquiry questionnaire), which we administer at unpredictable times. People are not informed when and how many times they pop up. The accuracy of recalled states is considered as people's awareness performance. After each condition, we administer the post-condition questionnaire. At the end of the study, we administer the post-test questionnaire.

### 5.2 Self-reported Questionnaires

Participants were asked to answer on-screen self-report questionnaires whenever they pop up. When a questionnaire is up, the interface blacks out, the team simulation pauses, and the robot stops.

Our demographics questionnaire asks people's age and biological sex, how often they play 3D video games and their game skill, how often they drive motor vehicles and their driving skill, and if they participated a robot study before. Our state inquiry questionnaire asks people to report the up-to-date team robot states, the confidence level for their choices, and write internal states on their own words. With the post-condition questionnaire, we collected people's level of nausea; sense of task speed; how much they felt the representation demanded their attention, was distracting, and helped them to maintain awareness; written comments; and NASA TLX [13]. With the post-study questionnaire, we collected their preference of representations, general thoughts on how their previous experience helped them in the task, and any comments.

Our quantitative report questions have a range of 0 (positive) to 20 (negative) inclusive for TLX reports and post-condition reports. The lower value on TLX reports means less task load and better performance. The awareness performance is normalized to the range of 0 to 100 inclusive, where 100 means the perfect (i.e., perfectly recalled all team robot states). To calculate the awareness score, we did not include a team robot's move and look states as our simulated robots are constantly moving and looking around.

We normalized the score for readers to quickly understand the numbers without scaling themselves, especially in the comparison between different types of state inquiry questionnaires (details in Pilot 5, Alternative Questionnaire).

### 5.3 Pilot 1, One Team Robot

The purpose of this initial pilot was to investigate overall feasibility of our teleoperation test bed, in addition to comparing the base case (the text representation) to other team robot state representations.

We have five within-subject conditions: base case (TEXT), animated icon with emoji (ANIMATED EMOJI), animated icon without emoji (ANIMATED ICON), non-animated icon (STATIC ICON), and non-animated icon with emoji (STATIC EMOJI).

Each condition is exactly 6 minutes and 30 seconds long excluding the time taking to answer the state inquire questionnaire. At a predefined timestamp, the questionnaire pops up without any notice to the participants. In each condition, it pops up three times. While the order of the simulated robots, questionnaire timestamps, and task locations fixed, the interface conditions were counterbalanced using an incomplete Latin Square.

At the beginning of each condition, the researcher reminds the participant of the condition and asks them if they have any questions.

We used our team robot state representations in colors (i.e., not grayscale version). Participants were asked to maintain their awareness of one team robot and answer our questionnaire (**Fig. 12.A**), which pops up three times at predefined timestamps per condition. We recruited 15 participants from the general university public.

#### 5.3.1 Qualitative Results

Five people preferred the text representations because they convey their meaning precisely, *"words are a much much more clear interpretations comparing with icons"* – P7 and *"I don't need to think [with TEXT]"* – P12.

Many (10 out of 15) liked icons or emojis as they are easy to understand and require comparably low mental demand, *"the text representation involved mentally deciphering the words and the color code associated with those words"* – P1 and *"I preferred the icons over the text because I found it easier to look at a picture representation than to read words while trying to complete a task"* – P9.

Many people (7 out of 10 people, excluding those who preferred the text) disliked facial expression add-ons. Main reason seems to be the confusion, *"I prefer the simple icons over the ones with character, because for me the ones with character was confusing me"* – P6 and *"the face is not clear and it is very big"* – P11. Three liked emojis mainly because they could get the overall situation, *"with character makes it easy to understand the overall situation of team*

*member" – P4* and *"because it [STATIC EMOJI] is easily to recognize and receive by specific characters, colors, than words description" – P5*.

Many people (7 out of 10) preferred the static representations over the animation. Negative comments toward animation were *"static can help in avoiding distractions" – P4* and *"the animated icons made the screen too busy so that I was feeling stressed and overwhelmed and couldn't focus on the task at hand as well" – P9*, and positive were *"I prefer the animation over static because animation grabbed my attention better than with just static. Also, you can see which state it is in in the corner of your eyes while trying to finish the given tasks" – P6*.

### 5.3.2 Quantitative Results

To investigate differences between the base case (TEXT) and the four representations (ANIMATED EMOJI, ANIMATED ICON, STATIC ICON, and STATIC EMOJI), we analyzed our results with repeated-measures one-way within-subject ANOVAs with planned contrasts comparing TEXT to the others. We found an effect on TLX performance report ($F_{4,56}=2.99$, $p<.05$). People reported that they performed ($F_{1,14}=4.78$, $p<.05$) better with STATIC EMOJI ($M=7.87$, $SD=5.14$) than TEXT ($M=5.53$, $SD=4.47$). We did not find any other effects.

To understand the impacts of animation and emotional information encoding, we analyzed our results with repeated-measures two-way within-subject ANOVAs (animated & static by with emoji & without) without the base case. We found an effect of reported TLX performance on EMOJI vs. NON-EMOJI. People reported that they performed ($F_{1,14}=5.92$, $p<.05$) better with EMOJI ($M=7.07$, $SD=4.06$) than NON-EMOJI ($M=5.00$, $SD=4.04$). We did not find any other effects nor interaction effects.

### 5.3.3 Lessons and Next Steps

We found that people think they performed differently with different team robot state indicators, while there was no statistical variation.

We thought that the difficulty of the awareness task was not high enough to see variation. Despite the average accuracy of awareness score which was 63.45 out of 100 with every representation, one participant perfectly recalled all states (i.e., scoring a perfect 100). To confirm this, we decided to increase the number of team robots in the next pilot.

As some people mentioned the benefits of color coding, we also planned to explore how color coding impacts people's awareness.

## 5.4 Pilot 2, Impact of the Number of Team Robots

To understand the impact of the number of team robots, we conducted a pilot study with two team robots. This study follows the same procedure as the first, has the same five conditions, and uses the same state inquiry questionnaire (**Fig. 12.A**).

We recruited 5 participants and compared new results to the first study's results to understand the impact of the number of team robots.

### 5.4.1 Results

We received a complaint regarding the task difficulty, *"there are too many things to look out for at the same time" – P21*, and a praise of using the color coding, *"same color represents same status is very good" – P23*.

One preferred the text representation, *"it is easy to remember while answering the questionnaire" – P21*. One preferred ANIMATED ICON, *"it was easier to interpret without focusing too much" – P24*. Three liked STATIC ICON, *"icon is easy to understand and simple and static can make me easy to find the moving and looking around information" – P23*.

### 5.4.2 Lessons and Next Steps

People mentioned verbally or through comments, that they were struggling to recall the states of the two team robots. We thought that our results should show variations on the performance of our awareness task due to the increased difficulty; however, we did not find a big variation.

While analyzing comments, we realized that question options in our state inquiry questionnaire are written in words, which may have given the text case an advantage and the other cases a disadvantage, due to mental conversion (e.g., for icon representations: from icon to text). We decided to update our state inquiry questionnaire for Pilot 4, Matching Options in the State Inquiry Questionnaire.

## 5.5 Pilot 3, Impact of Color Coding

To understand the impact of color coding, we recruited 5 people and tested our grayscale version with one team robot.

Every procedure including the state inquiry questionnaire was the same with the first. We compared new results to the first study's results.

### 5.5.1 Results

Overall, similar themes to the first study are emerged. Two liked the text team robot state representation, *"the words allowed me to understand the icons more quickly" – P28*. Three liked icons, *"because it is less demanding on the visual processing,*

**Fig. 12.A. text options**

**Fig. 12.B. alternative**

**Fig. 12.C. matching options**

**Fig. 12** Out (A) text, (B) matching option, and (C) alternative on-screen team robot state inquiry questionnaires. We did not include color coding in the questionnaire. When the questionnaire pops up, the teleoperation interface blacks out its screen, stops all processes, and waits for the questionnaire to be completed.

*has little to no distraction, gives room for better mental representations for the individual, which is absent in textual forms"* – *P30*. One pointed the emoji placement, *"emoji was hiding the movement of the robot"* – *P29*. One suggested an additional communication channel *"sound alert approach is better I think"* – *P26*.

### 5.5.2 Lessons and Next Steps

Our participants were only exposed to the grayscale version. Despite that, their comments were similar to the first study. Hence, we decided not to continue our studies with the grayscale versions.

Additionally, a person suggested different communication channel, the audio alert. We take this as an encouraging comment, because we already have an extra communication channel (color coding).

### 5.6 Pilot 4, Matching Options in the State Inquiry Questionnaire

Our original state inquiry questionnaire has text options for multiple choices (**Fig. 12.A**). That is, participants may have higher workload with icons and emojis due to the conversion to text when answering our questionnaire. Therefore, we made a new questionnaire with options that match to visual representation (**Fig. 12.C**). New questionnaire contains the grayscale version of the options (e.g., a static icon has a grayscale icon as an option).

During the training session, we made sure that we introduce the questionnaire to participants. We also made sure that the participants understand the questionnaire before proceeding to the next phase.

We recruited 10 participants to test the matching option questionnaire. The study procedure was the same to the Pilot 3, Impact of Color Coding.

### 5.6.1 Qualitative Results

Overall, similar themes to the first study continued to emerge with the results. People (5 out of 10) preferred the text team robot state representations because *"ICON is hard to memorize. I honestly prefer the color not the texture itself"* – *P34* and *"it is much more clearer in sending the message"* – *P35*. Others preferred icons *"because its easy for me to see and does not distract me from the task"* – *P31*.

A complaint regarding the number of states to remember raised again, *"two team members are too much. I could remember only one of them"* – *P40*, and message icon, *"I also feel that showing the 4th green circle [for the message state] is much better because it [sic.] there is no showing [visual indicator for the level of the message state] when it is normal, it is hard to remember"* – *P34*.

### 5.6.2 Quantitative Results

To investigate overall differences between the base case and other representations, we analyzed our results with repeated-measures one-way within-subject ANOVAs with planned contrasts comparing *TEXT* to others; we did not find any statistically significant effect.

To understand the impacts of animation and emoji, we analyzed our results with repeated-measures two-way within-subject ANOVAs without the base case. We found an effect on TLX effort report in *ANIMATED* vs. *STATIC*. People thought that ($F_{1,9}=7.57$, $p<.05$) *ANIMATED* ($M=15.05$, $SD=3.85$) requires more effort than *STATIC* ($M=13.75$, $SD=4.22$). We found an interaction effect on the TLX temporal report ($F_{1,9}=9.45$, $p<.05$): *ICON* (*STATIC*: $M=12.00$, $SD=4.00$, *ANIMATED*: $M=10.00$, $SD=4.67$) and *EMOJI* (*STATIC*: $M=9.70$, $SD=6.02$, *ANIMATED*: $M=11.60$, $SD=5.91$).

### 5.6.3 Lessons and Next Steps

We did not find a large variation on people's recall performance with different indicators. In addition, the update of our measurement does not impact people's performance. However, we gained a hint that the number of things to remember may decrease people's task performance and increase their cognitive load. Therefore, the questionnaire does not impact the performance significantly.

We wondered how the question style impacts people's performance. If we ask a fewer things at once, would this increase people's performance on recalling team robot states? To answer this question, we prepared another state inquiry questionnaire.

### 5.7 Pilot 5, Alternative Questionnaire

In addition to the matching option questionnaire which has a lot of information to recall for the participant at once, we developed a short version of the state inquiry questionnaire, called the *alternative state inquiry questionnaire* (**Fig. 12.B**). The alternative questionnaire is to solely check the difficulty of the monitoring task by asking two simple questions: regarding one robot (between two robots), pick the current value of one given state (among connectivity, battery level, damage, and message), and answer if the robot's states are generally positive or negative.

We removed the questions regarding the team robot's move states (i.e., moving or not moving, and looking around or not looking around) since we did not include this in our analyses in our earlier pilots.

The experiment mostly follows the previous study procedure. However, we removed the animation conditions as we were exploring the questionnaire specifically. Thus, we have three conditions: *TEXT*, *STATIC ICON*, and *STATIC EMOJI*. Unlike other questionnaires, the alternative questionnaire pops up eight times per condition. The recall performance is normalized to the range of 0 to 100.

Each condition is exactly 10 minutes long excluding the time taking to answer the state inquire questionnaire. While the orders of simulated robots, questionnaire timestamps, and task locations fixed, the interface conditions were counterbalanced.

To test the alternative questionnaire, we recruited 6 people. We include the existing data in our analysis to compare among our variants. Thus, we have 25 participants for the text option, 10 for the matching option, and 6 for the alternative questionnaire.

### 5.7.1 Quantitative Results

We analyzed our results on accuracy of recalling team robot states using various questionnaires with repeated-measures one-way between-subject ANOVAs. We found an effect on team robot state report accuracy ($F_{2,18}=14.37$, $p<.05$). People could report team robot states more accurately with the text questionnaire ($M=53.61$, $SE=4.66$) than the matching questionnaire ($M=41.67$, $SE=3.29$) and the alternative questionnaire ($M=20.83$, $SE=4.25$).

We conducted one-sample two-tailed t-test to check people's response to random chance (1/3 for each team robot state, as each state recall question has three options) with each questionnaire. We found that people's task performance with the matching option questionnaire in *TEXT* representation is close to random ($t_9=1.34$, $p>.05$). We found that people performed worse than random chance with the alternative questionnaire in two cases: *TEXT* ($t_5=-3.34$, $p<.05$) and *EMOJI* ($t_5=-3.09$, $p<.05$).

### 5.7.2 Lessons

We found that asking about a subset of possible states reduces the accuracy dramatically. People's performance was worse than a random chance. This leads us to think that the way of asking questions can be an important factor for measuring people's performance of a recall task.

## 6 Our Guidelines for Graphically Representing Robot Team Member States

Overall, all representations help people to maintain awareness of team robot states to a certain degree. From our findings, we created design guidelines.

### 6.1 Design Guidelines

*text is a viable candidate* – short one- or two-word text state representations performed well enough among the other, more icon representations. While one may assume that text is slow to read, perhaps with short text it can become similar to an icon and easily recognizable (iconification).

*people feel icons are easier* – more than half of the participants reported that they prefer icons and emojis, despite possible iconification of the text. According to their comments, they felt icons were easier to understand. Consider using icons in cases where people's perception of workload is important.

*anthropomorphic representations may not be clear* – while some enjoyed the faces, many participants reported that the emotional encoding information were not clear. While our goal was to provide high-level insight, perhaps this less-accurate representation may confuse people.

*animation: balance distraction with attention grabbing* – participants reported both sides of this regarding animated indicators. Some found that it is distracting while others found that it grabbed their attention. This supports prior work [28], and designers should leverage animation when needed to.

*color is good to show the level of robotic states* – according to our participants, color coding helped them to maintain their awareness of team robot states, because the color distinguishes the level of states (severity or urgency). We recommend teleoperation interface designers to use color coding for the level of robotic states if applicable.

## 7 Discussion & Limitations & Future Work

Our text team robot representation performed better than we anticipated. One explanation can be iconification. Our text state has color and two simple words. This may become an icon in participants' minds instead of text after a while. We propose to explore the text iconification further in the future.

People's thought on their awareness performance does not correlate to their actual performance with different team robot state indicators. This can be a good hint for interface designers – they may be able to alter user experience of robot state interfaces by changing information representations and graphical visualization parameters.

We found that some people liked our animated indicators and did not think they are distracting. While continuous animation may not always be necessary, using the animation at the appropriate moment such as transition between states can draw people's attention to help people to be aware of changes in team robot states.

Retrospectively, we realized that embedding a message into the emoji may hurt the recall task performance because people must extract the state from the face (i.e., an extra step to understand the state). This limitation on designing emojis and exploration of different emoji designs is future work.

In the evaluation, we did not include any actions for participants to take using the team robot state information. This may impact their motivation to maintain awareness of team robot states during the study. We leave the investigation of the relationship between active tasks using the team robot state information and people's overall awareness performance in the study as a future work.

We used a self-report questionnaire, NASA TLX, to measure people's task load including mental demand. To precisely measure one's cognitive load, we could use people's reaction time [8,37] or pupil diameter [10,16]. We are not sure how these dependent measures may result in some interesting contrasts between groups. We leave this as one of our future works.

In this work, we abstracted each robotic state into three different levels. Because of the exploratory nature of this work, we did not include any other abstractions; however, different levels of abstraction might have an impact to people's awareness and cognitive load. For example, two levels (okay or not okay) would add small cognitive load compared to three or five levels. In addition, how to visualize different levels of robotic states is questionable. We leave these questions to be answered in future work.

In our scenario, we assumed that an operator is directly teleoperating a robot while maintaining team members. With novel techniques, many robots become semi- or fully autonomous. This brings new challenges in designing teleoperation interfaces. For example, fully autonomous robots would allow the operator to only maintain high-level goals (e.g., planning navigation path for each robot or planning how to carry out victims). Some previous work (e.g., [17]) touched on the human-robot communication. However, in near future, we will need extensive explorations on how to design effective multi-robot teleoperation interfaces and how to represent robotic states of multiple robots in the interfaces.

From participants' comments, we can see that people have different preferences regarding our state representations (e.g., some like text representation and others like icon representations). One way to satisfy different preference is to provide customization tools in the interface. In this work, we provided the design guideline, which is mainly for teleoperation interface designers. However, an operator can have the design guideline and make an informed decision when they customize their interface. We leave the work of exploring the benefits and the impact of customization as a future work.

We found that asking a subset of robot states reduces participant accuracy. In addition, the matching options in the questionnaire do not seem to help people answer the team robot states accurately either. However, the reason for this is not clear. People may have a hard time because, for instance, it is simply hard to recall two team robots' states, robotic states were complicated, there were too many states for each robot, or the artistic style of our assets are noisy. We leave an exploration on question styles for measuring people's performance as a future work.

# 8 Conclusion

We presented our prototypes with different information representations and graphical visualization parameters. We developed a generally applicable team teleoperation scenario test bed for ours and future explorations of teleoperation interface techniques. We conducted a set of iterative exploratory studies, reported details including people's implications of our team robot state indicators, and presented a set of guidelines for graphically representing robot team member's states with our results and thought processes. We found that graphical interface elements may impact people's perception of their task performance without impacting actual awareness. In addition, we found that an inquiring method can greatly impact the accuracy. We suggest in-depth future explorations to further investigate our results, such as investigations into issues regarding questionnaires, and how user interface designs affect cognitively taxing team teleoperation scenarios.